\documentclass[preprint,a4,epsf]{revtex4}
\usepackage{graphicx,graphics} 

\def\s{\sigma}

\def\up{\uparrow}
\def\dd{\downarrow}
\def\las{\langle}
\def\ras{\rangle}

\def\nn{\nonumber}

\def\kp{{\bf k_\parallel}}
\begin{document}
\title{The effects of Mn concentration on spin-polarized transport through
ZnSe/ZnMnSe/ZnSe heterostructures}
\author{Alireza Saffarzadeh}
\altaffiliation[Corresponding author. ]{E-mail:
a-saffar@tehran.pnu.ac.ir} \affiliation{Department of Physics,
Tehran Payame Noor University, Fallahpour St., Nejatollahi St.,
Tehran, Iran}
\date{\today}

\begin{abstract}
We have studied the effects of Mn concentration on the ballistic
spin-polarized transport through diluted magnetic semiconductor
heterostructures with a single paramagnetic layer. Using a fitted
function for zero-field conduction band offset based on the
experimental data, we found that the spin current densities
strongly depend on the Mn concentration. The magnitude as well as
the sign of the electron-spin polarization and the tunnel
magnetoresistance can be tuned by varying the Mn concentration,
the width of the paramagnetic layer, and the external magnetic
field. By an appropriate choice of the Mn concentration and the
width of the paramagnetic layer, the degree of spin polarization
for the output current can reach 100\% and the device can be used
as a spin filter.
\end{abstract}

\maketitle

\newpage
\section{Introduction}
Recent advances in controlling the spin-polarized transport in
metallic magnetic structures, and successful applications of such
systems in electronic devices \cite{Prinz,Wolf} renewed worldwide
interest in diluted magnetic semiconductor (DMS) structures.
These materials are a promising component for the new spin-based
information technology, in which the spin degree of freedom of
the electron can be utilized to sense, store, process and
transfer information. II-VI DMSs \cite{Schmidt1,Fur}, are a class
of materials that has received much attention in recent years due
to their interesting physical properties and the potential
applications in integrated magneto-optoelectronic devices
\cite{Ohno}. They are known to be good candidates for effective
spin injection into a nonmagnetic semiconductor (NMS) because
their spin polarization is nearly 100\% and their conductivity is
comparable to that of typical NMS. Among different types of II-VI
DMSs, (Zn,Mn)Se is a very promising one for spin injection, which
has been previously used for spin injection experiments into GaAs
\cite{Fiederling}, and ZnSe \cite{Schmidt2,Slobodskyy,Syed}.

All A$^\mathrm{II}_{1-x}$Mn$_x$B$^\mathrm{VI}$ alloys are
direct-gap semiconductors like their
A$^\mathrm{II}$B$^\mathrm{VI}$ parent materials. In the presence
of a magnetic field, the band edges in these materials undergo a
huge spin splitting due to the $\emph sp-d$ exchange between
carriers and localized magnetic ions \cite{Fur}, while the
splitting in the nonmagnetic layers is much smaller. When the
large Zeeman splitting in the magnetic layers overcomes the band
offsets in both conduction and valance bands, the magnetic layers
act as barriers for electrons and holes in the spin-up state, and
as quantum wells for the spin-down state.

In recent years, the spin-polarized transport in II-VI DMS
heterostructures has been investigated theoretically. Based on a
quantum theory and the free-electron approximation, Egues
\cite{Egues}, Guo \emph{et al.} \cite{Y-Guo}, and Chang and
Peeters \cite{Ch-Pe} studied spin filtering in
ZnSe/Zn$_{1-x}$Mn$_x$Se/ZnSe heterostructures in the ballistic
region. The results showed a strong suppression for one of the
spin components of the current density with increasing the
external magnetic field. Also, Zhai \emph{et al.} \cite{Zhai}
investigated the effects of conduction band offset on the spin
transport in such heterostructures and showed that the positive
zero-field band offset can drastically increase the spin
polarization. However, they have not considered the effects of Mn
concentration and the width of the paramagnetic layer on the
spin-polarized transport. Thus other aspects of these
heterostructures remain to be explained.

In the present work, we study theoretically the dependence of
spin-polarized transport on Mn concentration in ZnSe/ZnMnSe/ZnSe
structures. The ZnSe layers are considered as emitter and
collector attached to external leads. We assume that the carrier
wave vector parallel to the interfaces and the carrier spin are
conserved in the tunneling process through the whole system.
These assumptions can be well justified for interfaces between
materials whose lattice constants are nearly equal; and when the
sample dimensions are much smaller than the spin coherence
length. Using a quantum theory, we study the effects of Mn
concentration and the width of the paramagnetic layer on
spin-dependent current densities, the degree of spin
polarization, and the magnetoresistance ratio.

In section 2, we present a fitted function for zero-field
conduction band offset and describe the model. Then, the spin
current densities, the electron-spin polarization, and the
magnetoresistance ratio for ZnSe/ZnMnSe/ZnSe heterostructures are
formulated. In section 3, the numerical results for the above
quantities are discussed in terms of the Mn concentration, the
width of the paramagnetic layer, and the applied voltage. The
results of this work are summarized in section 4.

\section{Model and formalism}
Consider a spin unpolarized electron current injected into
ZnSe/Zn$_{1-x}$Mn$_x$Se/ZnSe structures in the presence of
magnetic and electric fields along the growth direction (taken as
$z$ axis). In Mn-based DMS systems, the conduction electrons that
contribute to the electric currents, interact with the 3d$^5$
electrons of the Mn ions with spin $S=\frac{5}{2}$ via the sp-d
exchange interaction. Due to the sp-d exchange interaction, the
external magnetic field gives rise to the spin splitting of the
conduction band states in the Zn$_{1-x}$Mn$_x$Se layer.
Therefore, the injected electrons see a spin-dependent potential.
In the framework of the parabolic-band effective mass
approximation, the one-electron Hamiltonian of such system can be
written as
\begin{equation}\label{H}
H=\frac{1}{2m^*}({\bf P}+e{\bf
A})^2+V_s+V_x(z)+V_{\s_z}(z)-\frac{eV_az}{d} \ ,
\end{equation}
where the electron effective mass $m^*$ is assumed to be
identical in all the layers, and the vector potential is taken as
${\bf A}=(0,Bx,0)$. Here,
$V_s=\frac{1}{2}g_s\mu_B{\bf\s}\cdot{\bf B}$ describes the Zeeman
splitting of the conduction electrons, where ${\bf\s}$ is the
conventional Pauli spin operator; $V_x(z)$ is the heterostructure
potential or the conduction band offset in the absence of a
magnetic field, which depends on the Mn concentration $x$ and is
the difference between the conduction band edge of the
Zn$_{1-x}$Mn$_x$Se layer and that of the ZnSe layer;
$V_{\s_z}(z)$ is the sp-d exchange interaction between the spin
of injected electron and the spin of Mn ions and can be
calculated within the mean field approximation. Hence, the sum of
the last two terms can be written as
\begin{eqnarray}\label{U}
V_x(z)+V_{\s_z}(z)&=&[(1/2)\Delta E(x)-N_0\alpha\s_zx_{eff}\las
S_{z}\ras] \Theta(z)\Theta(d-z)\  ,
\end{eqnarray}
where
\begin{eqnarray}\label{Veff}
\Delta E(x)=-0.63x+22x^2-195x^3+645x^4 \ .
\end{eqnarray}
Here, $\Delta E(x)(=E_g(x)-E_g(0))$ is the sum of the conduction
and valance band offset under zero magnetic field, when the real
Mn concentration is $x$. In Fig. 1, we have shown the Mn
concentration dependence of the energy gap $E_g(x)$ in
Zn$_{1-x}$Mn$_x$Se layer. It is clear that the band gap of
Zn$_{1-x}$Mn$_x$Se varies anomalously with $x$; it shows a
minimum at $x\sim 0.02$ and increases linearly with $x$ for $x\geq
0.05$. This anomalous behavior of Zn$_{1-x}$Mn$_x$Se which called
band-gap bowing, probably refers to the sp-d exchange interaction
\cite{Fur}. In deriving $\Delta E(x)$, we have used of the fitted
curve of $E_g(x)$. We have approximated $\Delta E(x)/2$ as the
zero-field conduction band offset. This function does not depend
on spin, and for $x\leq 0.043$ behaves as a quantum well, and as
a potential barrier for $x\geq 0.043$.

In Eq. (\ref{U}), $N_0\alpha$ is the electronic sp-d exchange
constant, $\s_z=\pm1/2$ (or $\up,\dd$) are the electron spin
components along the magnetic field, $x_{eff}=x(1-x)^{12}$ is the
effective Mn concentration used to account for antiferromagnetic
Mn-Mn coupling, and $\Theta(z)$ is the step function. $\las
S_z\ras$ is the thermal average of $z$th component of Mn$^{2+}$
spin in the paramagnetic layer which is given by the modified
$\frac{5}{2}$ Brillouin function $SB_S[5\mu_BB/k_B(T+T_0)]$,
where $T_0$ is an empirical parameter representing
antiferromagnetic interactions between the Mn ions \cite{Awsch}.
The last term in Eq. (\ref{H}) denotes the effect of an applied
voltage $V_a$ along the $z$ axis on the system, and $d$ is the
width of the paramagnetic layer. It is important to note that,
$d$ is much smaller than the spin coherence length in the
semiconductors. Therefore, we have neglected the effects of
spin-flip processes in the Hamiltonian of the system.

In the absence of any kind of scattering center for the electrons,
the motion along the $z$-axis is decoupled from that of the
$x$-$y$ plane. Therefore, in the presence of magnetic field $B$,
the in-plane motion is quantized in Landau levels with energies
$E_n=(n+1/2)\hbar\omega_c$, where $n=0,1,2,\cdots$ and
$\omega_c=eB/m^*$. In such case, the motion of electrons along
the $z$ axis can be reduced to the following one-dimensional (1D)
Schr\"odinger equation
\begin{equation}\label{HH}
-\frac{\hbar^2}{2m^*}\frac{d^2\psi_{\s_z}(z)}{dz^2}+U_{\s_z}(z,B)\psi_{\s_z}(z)
=E_z\psi_{\s_z}(z) \ ,
\end{equation}
where $E_z$ is the longitudinal energy of electrons and
$U_{\s_z}(z,B)=V_s+V_x(z)+V_{\s_z}(z)-eV_az/d$ is the effective
potential seen by a traverse electron, which includes the effects
of spin, conduction band offset, and external magnetic and
electric fields. The general solution to the above Schr\"odinger
equation is as follows:
\begin{equation}\label{psi}
\psi_{j\s_z}(z)=\left\{\begin{array}{cc}
A_{1\s_z}e^{ik_{1\s_z}z}+B_{1\s_z}e^{-ik_{1\s_z}z}, & z<0 ,\\
A_{2\s_z}{\rm Ai}[\rho_{\s_z}(z)]+B_{2\s_z}{\rm Bi}[\rho_{\s_z}(z)] , & 0<z<d ,\\
A_{3\s_z}e^{ik_{3\s_z}z}+B_{3\s}e^{-ik_{3\s_z}z}, & z>d ,\\
\end{array}\right.
\end{equation}
where $k_{1\s_z}=\sqrt{2m^*(E_z-V_s)}/\hbar$,
$k_{3\s_z}=\sqrt{2m^*(E_z-V_s+eV_a)}/\hbar$; Ai[$\rho_{\s_z}(z)$]
and Bi[$\rho_{\s_z}(z)$] are Airy functions with
$\rho_{\s_z}(z)=[d/(eV_a\lambda)](E_z-U_{\s_z})$, and
$\lambda=[-\hbar^2d/(2m^*eV_a)]^{1/3}$; $A_{j\s_z}$ and
$B_{j\s_z}$ (with $j$=1-3) are constants which can be obtained by
matching the wave functions and their derivatives at the
interfaces of Zn$_{1-x}$Mn$_x$Se and ZnSe layers. The matching
results in a system of equations, which can be represented in a
matrix form \cite{Allen},
\begin{eqnarray}
\left(\begin{array}{cc}
A_{1\s_z}\\B_{1\s_z}
\end{array}\right)
&=&M_1^{-1}(0)M_2(0)M_2^{-1}(d)M_3(d)\left(\begin{array}{cc}
A_{3\s_z}\\B_{3\s_z}
\end{array}\right)\nn \\
&=&M_{total}\left(\begin{array}{cc} A_{3\s_z}\\B_{3\s_z}
\end{array}\right) .
\end{eqnarray}
Here, $M_{total}$ is the transfer matrix that connects the
incidence and transmission amplitudes, and
\begin{equation}
M_j(z_i)=\left(\begin{array}{cc}
\psi_j^+(z)&\psi_j^-(z)\\
\frac{d\psi_j^+(z)}{dz}&\frac{d\psi_j^-(z)}{dz}
\end{array}\right)_{z=z_i}  \\,
\end{equation}
where, $\psi_j^+(z)$ and $\psi_j^-(z)$ are, respectively, the
first and second term of the wave functions in each layer, without
considering their coefficients. Therefore, the transmission
coefficient of the spin $\s_z$ electron, which is defined as the
ratio of the transmitted flux in the collector to the incident
flux in the emitter, can be written as
\begin{equation}
T_{\s_z}(E_z,B,V_a)=\frac{k_{3\s_z}}{k_{1\s_z}}
\left|\frac{1}{M_{total}^{11}}\right|^2 \ ,
\end{equation}
where $M_{total}^{11}$ is the (1,1) element of the matrix
$M_{total}$. The spin-dependent current density can be determined
by
\begin{eqnarray}\label{J}
J_{\s_z}(B)&=&J_0B\sum_{n=0}^\infty\int_{0}^{+\infty}T_{\s_z}(E_z,B,V_a)\nonumber\\
&&\times\{f[E_z+(n+\frac{1}{2})\hbar\omega_c+V_s]
-f[E_z+(n+\frac{1}{2})\hbar\omega_c+V_s+eV_a]\}dE_z \ ,
\end{eqnarray}
where $J_0=e^2/4\pi^2\hbar^2$, $f(E)=1/\{1+\exp[(E-E_F)/k_BT]\}$
is the Fermi-Dirac distribution function in which $k_B$ is the
Boltzmann constant, $T$ is the temperature, and $E_F$ denotes the
emitter Fermi energy.

The degree of spin polarization for electrons traversing the
heterostructure is defined by
\begin{equation}\label{p}
P=\frac{J_\dd(B)-J_\up(B)}{J_\dd(B)+J_\up(B)} \ ,
\end{equation}
where $J_\up$ $(J_\dd)$ is the spin-up (spin-down) current
density. On the other hand, in the absence of external magnetic
field, the Zeeman splitting of the conduction electrons $V_s$ and
the spin dependent potential $V_{\s_z}(z)$ are zero. In this
case, the effective potential reduces to
$U_{\s_z}(z,0)=V_x(z)-eV_az/d$ which is spin-independent. Thus,
by using $U_{\s_z}(z,0)$ and a procedure completely analogous to
the one used for the case of $B\neq 0$, one can obtain the
following formula for the total electric current density
\cite{Duke}
\begin{equation}\label{j}
J(0)=2\left(\frac{em^*k_BT}{4\pi^2\hbar^3}\right)\int_0^{+\infty}T_{\s_z}(E_z,0,V_a)
\ln\left\{\frac{1+\exp[(E_F-E_z)/k_BT]}{1+\exp[(E_F-E_z-eV_a)/k_BT]}\right\}
dE_z \ .
\end{equation}
The factor two is due to the degeneracy of the electron spin in
the case of $B=0$. Although the transverse momentum $\kp$ was not
appeared in the above equation, the effects of the summation over
$\kp$ have been considered in our calculations. Here, we mention
again that the effective mass is independent of layer. When
taking both the transverse motion and the layer-dependent
effective mass of the electron into account, the transmission
coefficient can have a pronounced dispersion in $\kp$ space
\cite{Vosk}. In this case, one cannot simply reduce the 3D
Schr\"odinger equation to the 1D equation and integrate the $\kp$
to obtain the current density, as we did here.

The linear conductances per unit area are given by $G(0)=J(0)/V_a$
and $G(B)=\sum_{\s_z}J_{\s_z}(B)/V_a$ for $B=0$ and $B\neq0$,
respectively. Therefore, the tunnel magnetoresistance (TMR) or
magnetoresistance ratio in such heterostructures can be described
quantitatively by
\begin{eqnarray}\label{tmr}
\mbox{TMR}&=&\frac{G(0)-G(B)}{G(B)} \nn \\
&=&\frac{J(0)}{J_\up(B)+J_\dd(B)}-1 \ .
\end{eqnarray}
In next section, we will present the numerical results for
$J_\s$, $P$ and TMR in terms of Mn concentration.

\section{Results and discussion}
In our numerical calculation we have taken the following values:
$m^*=0.16$ $m_e$ ($m_e$ is the mass of the free electron), $E_F=5$
meV, $B=4$ T, $g_s=1.1$, $T=2.2$ K, $T_0=1.7$ K, and
$N_0\alpha=-0.26$ eV \cite{Konig}. Fig. 2(a)-(d) show the
dependence of spin-up current densities on the Mn concentration
for several choices of $d$ and $V_a$. In all of the widths of the
paramagnetic layer, for a fixed value of Mn concentration, the
spin-up current density increases with the applied voltage. With
increasing the Mn concentration ($x>0.043$), the spin-up
electrons see a higher potential barrier, while with increasing
the width of the paramagnetic layer, these electrons see a thicker
potential barrier. Thus in both cases, the tunneling probability
for this group of electrons decreases and this leads to a
reduction of the current density. From the figures, we also find
that, for $d=$200 nm and $x>0.055$, the transmission for spin-up
electrons is completely suppressed. On the other hand, with
increasing the applied voltage, the potential barrier tilts and
the effective width of the barrier becomes narrower; therefore,
the spin-up current density increases, as shown in Fig. 2.

The dependence of spin-down current densities on the Mn
concentration is shown in Fig. 3(a)-(d) for several values of $d$
and $V_a$. For small widths of the paramagnetic layer, the
variations of the current density are relatively small with an
oscillatory behavior. These variations increase with the width
$d$. Qualitatively, the voltage dependence of spin-down current
density is similar to the spin-up one. This means that both spin
current densities increase with the applied voltage. We should
note that at $x=0$ the current densities are nearly
spin-independent, because $V_{\s_z}$ is zero and $V_s$ is very
small. With increasing $d$ and $x$, peaks are observed in the
spin-down current densities. The reason is that, the paramagnetic
layer behaves as a quantum well for spin-down electrons; thus, the
enhancement of the width of the paramagnetic layer, varies and
shifts the position of the resonant states formed in the well to
the lower energy region. This leads to the formation of peaks in
the spin-down current densities \cite{Saffar}.

In Fig. 4(a)-(d) we show the spin polarization for electrons
traversing the heterostructure as a function of $x$ for various
values $d$. According to the above discussion, for $x=0$ the
spin-up and spin-down current densities have equal values, hence
the spin polarization is zero. As the Mn concentration increases,
the spin-down current density first decreases, while the spin-up
one is approximately constant; therefore, the spin polarization is
negative. On further increasing the Mn concentration, the spin-up
current density decreases exponentially; however the spin-down
one can increase and when $J_\up$ vanishes, $J_\dd$ has non-zero
values. Consequently, the spin polarization becomes positive and
increases for these values of the Mn concentration. It is also
important to note that, for widths near 200 nm, the spin
polarization approaches 100\%, which shows that, the system acts
as a spin filter.

The magnetoresistance ratio of ZnSe/Zn$_{1-x}$Mn$_x$Se/ZnSe
heterostructures is another quantity which is sensitive to the Mn
concentration. In Fig. 5(a)-(d) the TMR is plotted as a function
of $x$. With increasing the width of Zn$_{1-x}$Mn$_x$Se layer, a
peak appears in the TMR curves. This peak is shifted slowly
towards lower values of the Mn concentration $x$, and its height
increases with the width. From the figures, it is clear that the
sign of TMR can be positive when $G(0)>G(B)$ or negative when
$G(0)<G(B)$. The reason of negative TMR can be understood by
considering the effects of Mn concentration on spin transport in
such heterostructures. With increasing the Mn content
($x>0.043$), the conduction band offset increases and the
paramagnetic layer acts as a barrier for both spin-up and
spin-down electrons in zero magnetic field. On applying a
magnetic field, however, the spin-down electrons see a quantum
well (when $V_x+V_{\s_z}<0$) and the current density for such
electrons increases. Hence, $G(0)<G(B)$ and TMR becomes negative.
The results also show that, for a fixed width of the paramagnetic
layer, the electron-spin polarization and the TMR curves do not
depend strongly on the applied voltage, as shown in Figs. 4 and 5.

Guo ${\it et~al.}$ \cite{Y-Guo}, studied the effects of
zero-field conduction band offset on spin transport with $V_x=-5,
0, +5$ meV and $x=0.05$. They found that, for $V_x=+5$ meV the
spin currents are highly polarized, while, for $V_x=-5$ meV the
spin polarization is very low. Our present results based on the
experimental data, however, indicate that the conduction band
offset depends on the Mn concentration. Thus, in order to
understand the correct description of the effects of the
conduction band offset on spin transport, the dependence of this
quantity on the Mn concentration was included. It is clear that
the width of the paramagnetic layer is also one of the main
factors in spin-polarization of the output current of the system.

Therefore, the obtained results clearly illustrate that the
current densities and hence the degree of spin polarization and
the TMR can be tuned by changing the Mn concentration and/or the
width of the paramagnetic layer.

\section{Summary}
In this paper, using the transfer matrix method and the
effective-mass approximation we investigated the ballistic
spin-polarized transport in ZnSe/Zn$_{1-x}$Mn$_x$Se/ZnSe
heterostructures. We examined the effects of Mn concentration,
the width of the paramagnetic layer, and the external magnetic and
electric fields on the spin current densities, the electron-spin
polarization and the TMR. The numerical results show that the
zero-field conduction band offset which varies with the Mn
concentration, plays an important role in the spin current
densities. The spin polarization and the TMR are not very
sensitive to the applied voltage. However, by adjusting the Mn
concentration and the width of the paramagnetic layer, the output
current exhibits a nearly 100\% spin polarization, and also the
sign of the TMR can be positive or negative. The presented results
may be helpful from a technological application point of view
such as the generation of spin-polarized injection electrons.

\newpage
\begin{figure}
\centering \resizebox{0.8\textwidth}{0.48\textheight}
{\includegraphics{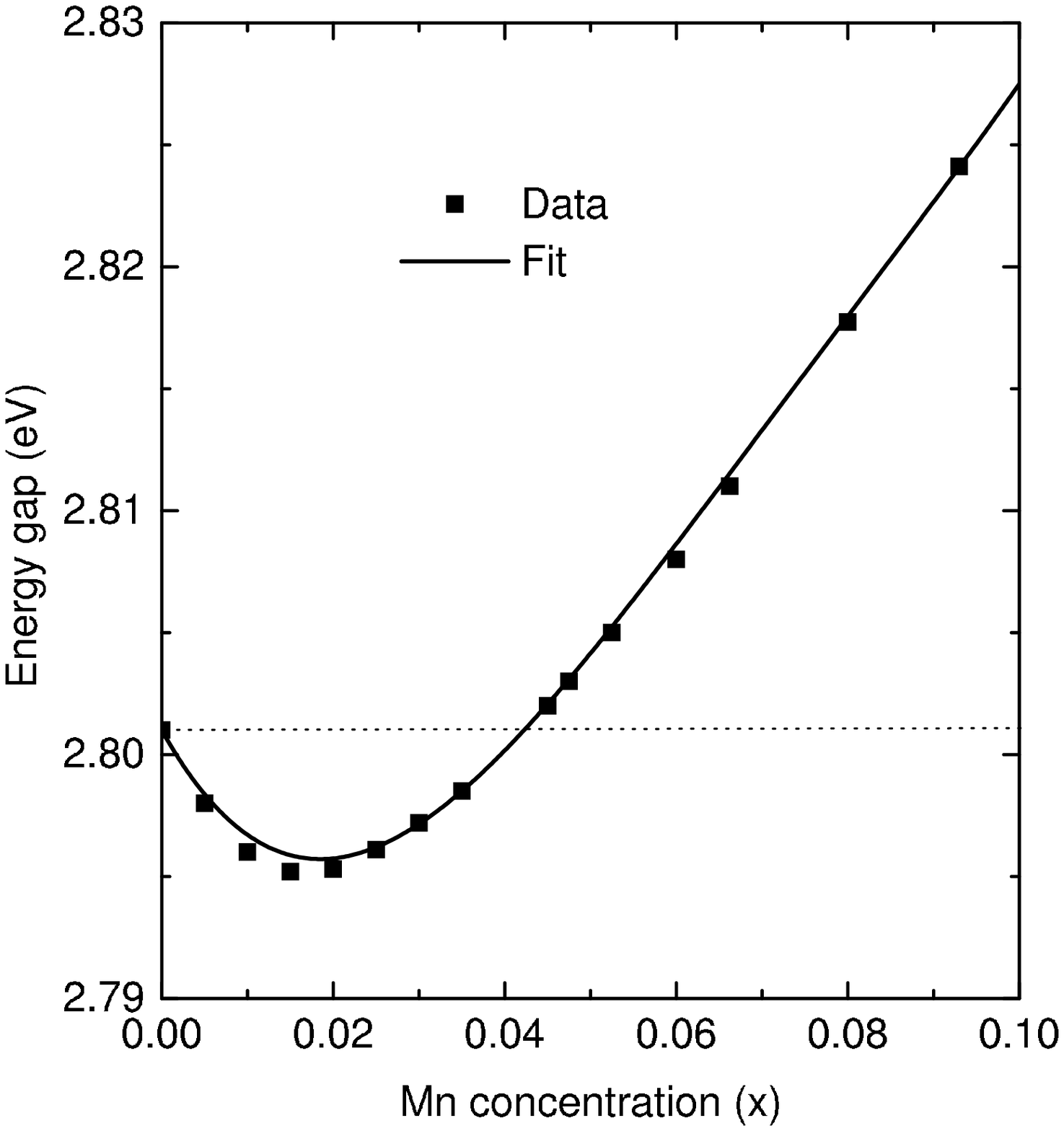}} \caption{Energy gap of
Zn$_{1-x}$Mn$_x$Se as a function of Mn concentration $x$ at
$T$=2.2 K. The experimental data (full squares) is taken from
\cite{Dai}. The solid curve is a fit to the data.}
\end{figure}
\newpage
\begin{figure}
\centering \resizebox{0.8\textwidth}{0.48\textheight}
{\includegraphics{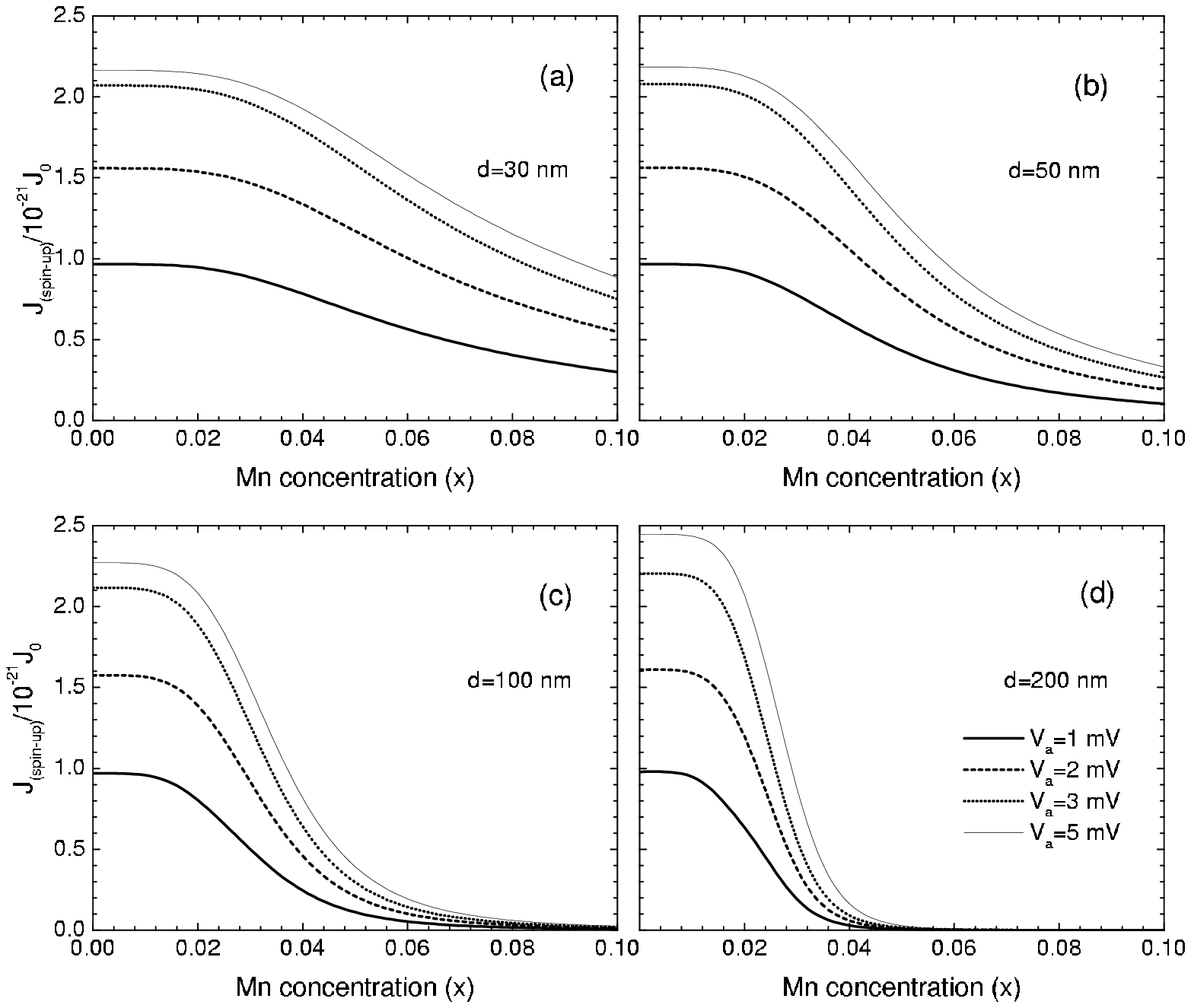}} \caption{Spin-up current densities
as a function of Mn concentration $x$ for different applied
voltages and widths of Zn$_{1-x}$Mn$_x$Se layer.}
\end{figure}
\newpage
\begin{figure}
\centering \resizebox{0.8\textwidth}{0.48\textheight}
{\includegraphics{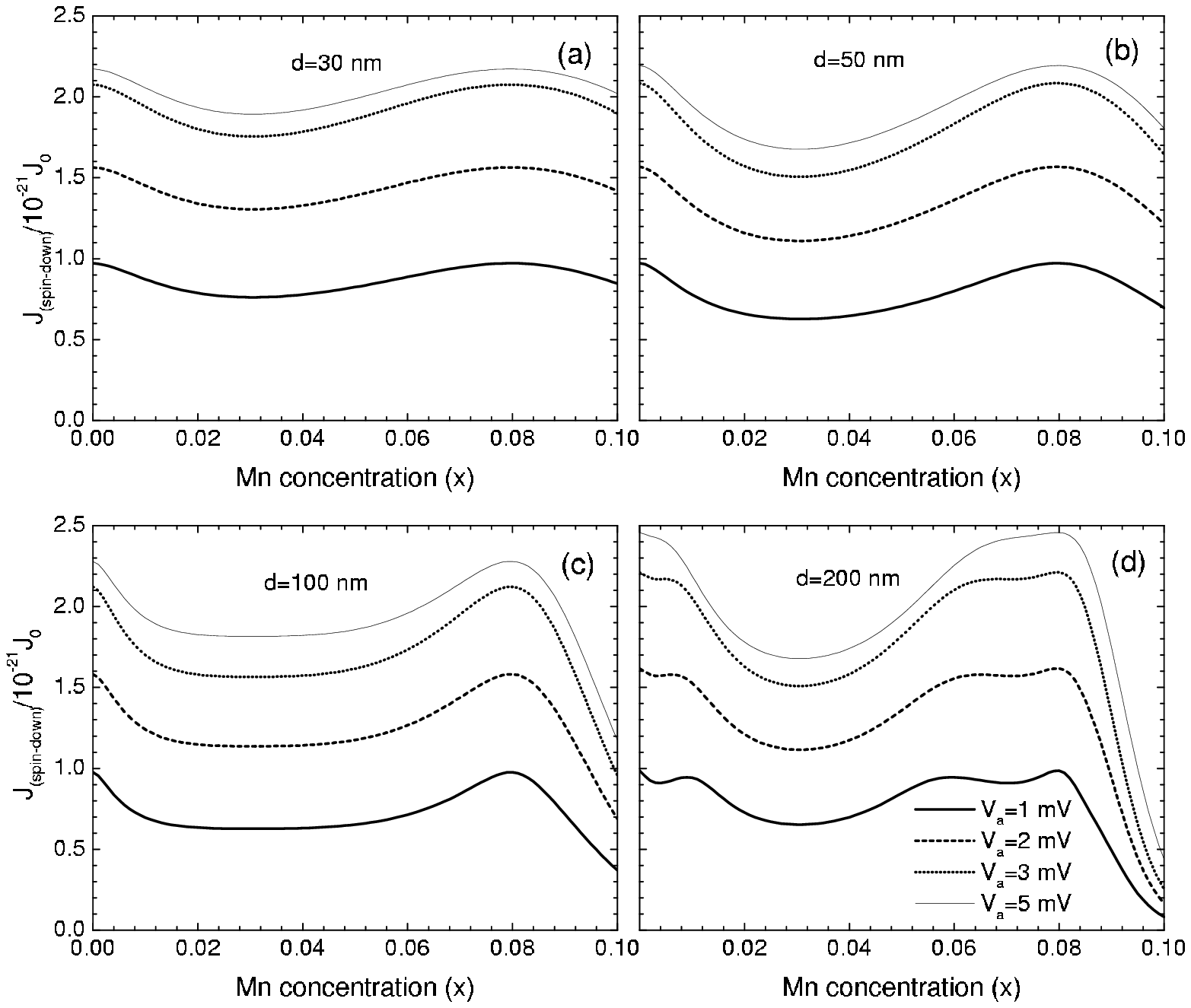}} \caption{Spin-down current densities
as a function of Mn concentration $x$ for different applied
voltages and widths of Zn$_{1-x}$Mn$_x$Se layer.}
\end{figure}
\newpage
\begin{figure}
\centering \resizebox{0.8\textwidth}{0.48\textheight}
{\includegraphics{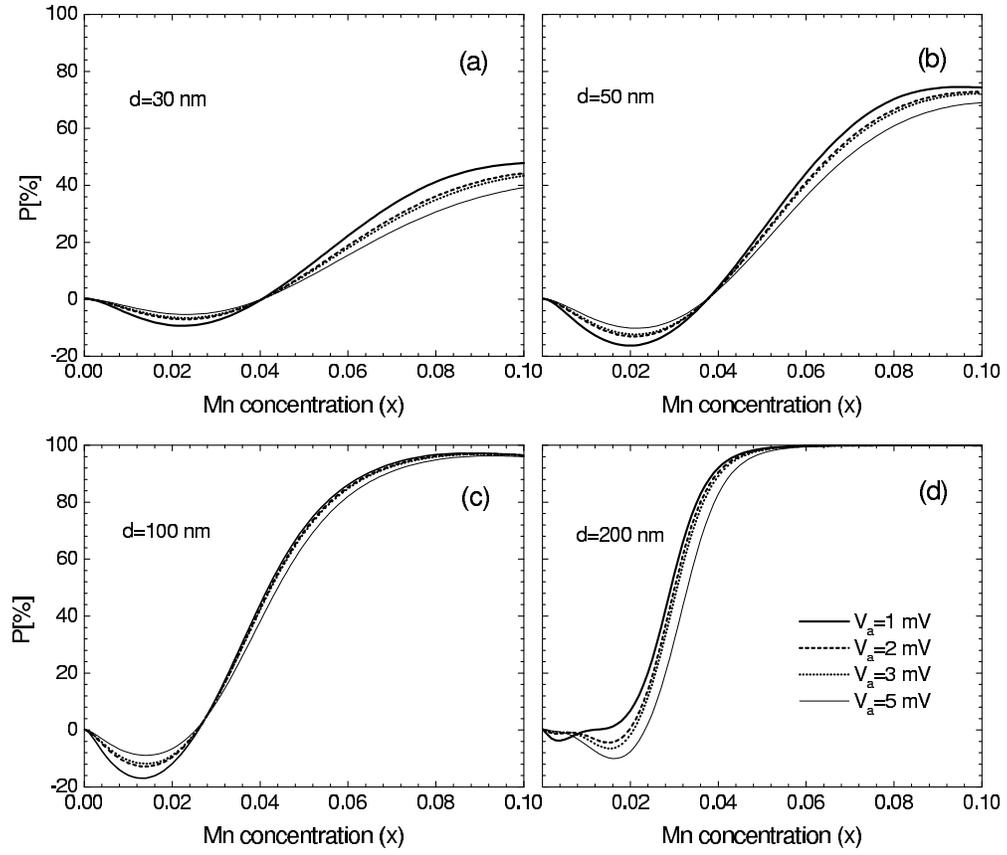}} \caption{Electron-spin polarization
$P$ as a function of Mn concentration $x$ for different applied
voltages and widths of Zn$_{1-x}$Mn$_x$Se layer.}
\end{figure}
\newpage
\begin{figure}
\centering \resizebox{0.8\textwidth}{0.48\textheight}
{\includegraphics{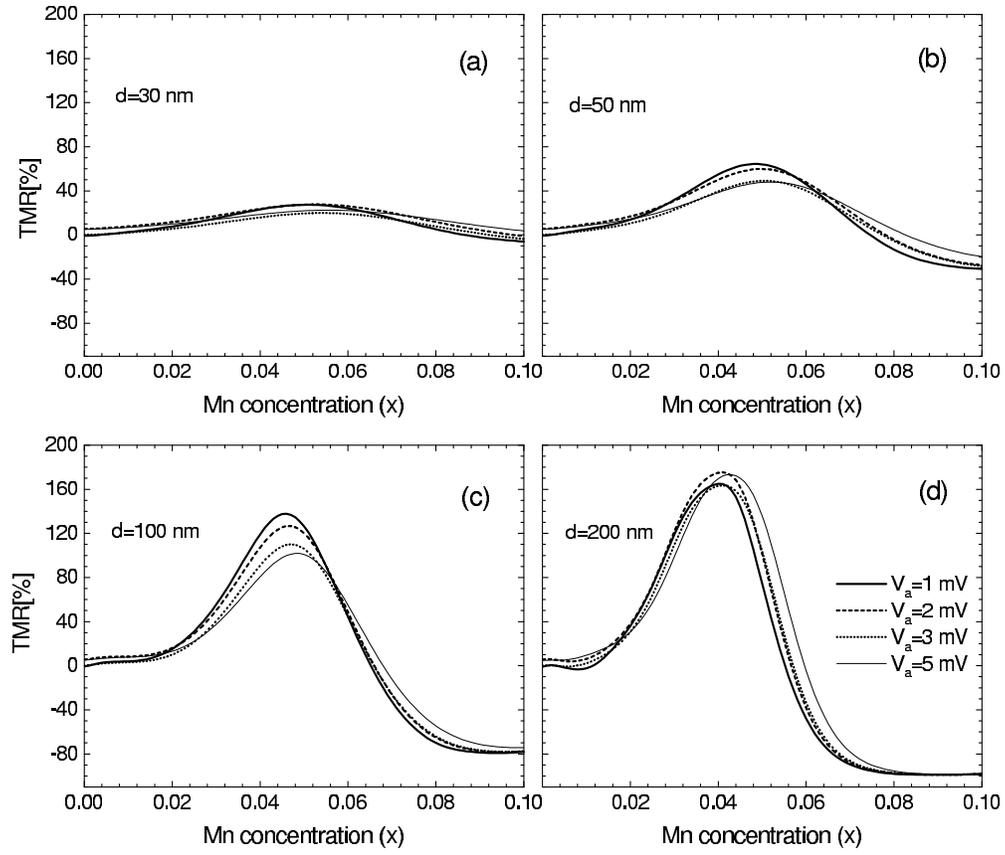}} \caption{TMR as a function of Mn
concentration $x$ for different applied voltages and widths of
Zn$_{1-x}$Mn$_x$Se layer.}
\end{figure}

\end{document}